\documentclass[12pt]{article}

\usepackage[totalheight = 23cm, totalwidth = 17cm]{geometry}
\usepackage{amssymb,amsmath,amsfonts,amsbsy,graphicx,axodraw}

\def\mathbi#1{\textbf{\em #1}}

\newcommand{\fnl}{{f_{\rm NL}}}
\newcommand{\nfnl}{{n_{f_{\rm NL}}}}

\newcommand{\calO}{{\cal O}}
\newcommand{\calP}{{\cal P}}
\newcommand{\calR}{{\cal R}}

\begin{document}

\begin{titlepage}

\begin{center}

\rightline{CERN-PH-TH/2012-249}

\vskip 1.5cm

\Huge{General formula for the running of local $\fnl$}

\vskip 1cm

\large{
Christian T. Byrnes$^{\star,\;\diamond}$\footnote{cbyrnes@cern.ch}
\hspace{0.2cm} and \hspace{0.2cm}
Jinn-Ouk Gong$^{\star}$\footnote{jinn-ouk.gong@cern.ch}
\\
\vspace{0.5cm}
{\em
$^{\star}$Theory Division, CERN, CH-1211 Gen\`eve 23, Switzerland
}
\\
\vspace{0.5cm}
{\em
$^{\diamond}$Astronomy Centre, University of Sussex, Brighton, BN1 9QH, UK
}}

\vskip 0.5cm

\today

\vskip 1.2cm

\end{center}

\begin{abstract}

We compute the scale dependence of $\fnl$ for models of multi-field inflation, allowing for an arbitrary field space metric. We show that, in addition to multi-field effects and self interactions, the curved field space metric provides another source of scale dependence, which arises from the field-space Riemann curvature tensor and its derivatives. The scale dependence may be detectable within the near future if the amplitude of $\fnl$ is not too far from the current observational bounds.

\end{abstract}

\end{titlepage}

\setcounter{page}{0}
\newpage
\setcounter{page}{1}

\section{Introduction}

Inflation~\cite{inflation}, the dominant paradigm of the early universe, successfully predicts the statistics of the primordial perturbations, within the observational limits seen today. However many fundamental questions remain, about the form of the inflation Lagrangian, how many fields were involved, etc. The upcoming large gains in sensitivity mean that there is a realistic chance to provide definite answers to some of them. In particular, non-Gaussianity contains far more information than the power spectrum. Non-Gaussianity is most commonly parametrised through the bispectrum and for a particular shape (e.g.~local or equilateral) all information is then contained in a single amplitude, $\fnl$~\cite{Komatsu:2001rj}. However, just as is the case for the power spectrum, it is rather natural for $\fnl$ to be scale dependent. This scale dependence is not strongly constrained and therefore could even be much stronger. Provided that the fiducial value of $\fnl$ is large enough, there is a realistic chance for the Planck satellite to simultaneously provide the first measurement of both $\fnl$ and its scale dependence parametrised by $\nfnl$ in the near future.

One of the aims of early universe physics is to test fundamental theories at far higher energy scales than can ever be reached by terrestrial experiment. High energy theories such as supersymmetry or string theory typically predict that there are multiple scalar fields, and that the kinetic term involves a non-trivial field metric determined by e.g. K\"ahler potential~\cite{Nilles:1983ge}. This field metric affects observable parameters but this topic has not been very extensively studied. Most previous work was restricted to the power spectrum, i.e.~to linear perturbations. In this case it was shown that the field metric gives rise to a new term in the spectral index which may easily be as large as the usual slow-roll terms \cite{Sasaki:1995aw,Nakamura:1996da,Gong:2002cx}. Only very recently the first detailed study of the bispectrum has been made, following the direction of~\cite{Gong:2011uw}, in the general case by Elliston et al.~\cite{Elliston:2012uy}. They provided covariant formulae for the bispectrum of the field fluctuations at horizon crossing and extended the $\delta N$ formalism~\cite{Sasaki:1995aw,Nakamura:1996da,Gong:2002cx,Starobinsky:1986fxa} to provide a generalisation of the well-known $\delta N$ result for the trivial field space metric~\cite{Lyth:2005fi} to a curved one. There have been previous studies of non-Gaussianity from a curved field space metric, but none so general or providing explicitly covariant results, see for example~\cite{curvedspaceNGprevious}.

Over the last few years, it has become clear that the scale dependence of local $\fnl$ is also a sensitive probe of early universe physics, and that observations are sensitive to it. Often the scale dependence, $\nfnl$, is of the same magnitude as the spectral index $n_\calR-1$, and in some models it may be much larger, for example the self-interacting curvaton scenario~\cite{intcurvaton}. Non-Gaussianity may also change rapidly over a short range of scales from being zero on large scales to large on smaller scales~\cite{Riotto:2010nh}. Scale dependence of equilateral $\fnl$ was first considered in~\cite{Chen:2005fe} and for the local model in~\cite{Byrnes:2008zy} (for the specific case of two-field hybrid inflation), with a more general formalism developed in~\cite{more-nfNL,Byrnes:2010ft}. Numerous studies of the scale dependence have been made for a large variety of models, for an incomplete selection see~\cite{scaledepfNL}. A general lesson is: any detection of primordial local non-Gaussianity is very valuable since it can rule out all single field models which predict a value of $\fnl$ in the squeezed limit proportional to the spectral index~\cite{singlefieldconsistency} (see however~\cite{Ganc:2012ae}), but for most models in which $\fnl$ can be large its amplitude can be tuned. Therefore a further observable, either the trispectrum or the scale dependence of $\fnl$, will still be required to help discriminate between the large number of models. In this paper we will always focus on the local model of non-Gaussianity.

Sefussati et al. have forecasted that for (local) non-Gaussianity, Planck could reach a sensitivity of $\sigma_{\nfnl}\sim0.1$ for a fiducial value of $\fnl=50$~\cite{Sefusatti:2009xu}. This is not much larger than the currently measured central value for the deviation of the spectral index from scale invariance, $n_\calR-1\sim-0.04$~\cite{Komatsu:2010fb}, and shows that there is a real possibility that Planck is able to measure both $\fnl$ and $\nfnl$ together, provided that the fiducial values are large enough. Numerous other forecasts have been made especially for surveys including large scale structure (LSS) data~\cite{runningfNL-LSS} and of course the tightest constraints may be expected to come from combining the cosmic microwave background and LSS data~\cite{Becker:2012yr}. This is a topical field and very recently the first real constraints on $\nfnl$ have been made in~\cite{Becker:2012je} (see also \cite{runningfNL-CMB}).

This article is structured as follows. In Section~\ref{sec:evolution} we study the evolution of the perturbations and rederive the spectral index of the power spectrum as a warm up. In Section~\ref{sec:bispectrum} we derive a general formula for the scale-dependence of $\fnl$, the main result of our paper, and we check our result by showing that it reduces to the known result in the case of a trivial field space metric. Finally we conclude in Section~\ref{sec:conclusions}.

\section{Evolution of field fluctuations}
\label{sec:evolution}

We study a period of inflation driven by an arbitrary number of scalar fields, labeled with indices $a,b,c,\cdots$, with Lagrangian density
\begin{equation}
\mathcal{L}=-\frac12 \gamma_{ab}g^{\mu\nu}\partial_{\mu}\phi^a\partial_{\nu}\phi^b-V\,, 
\end{equation}
where $g_{\mu\nu}$ is the usual space-time metric, and $\gamma_{ab}$ is the field metric, which is a function of the field values. In the case of a trivial field space metric, $\gamma_{ab}=\delta_{ab}$. The potential $V$ is also an arbitrary function of field values, except that we assume that it is sufficiently flat so that we may use the slow-roll approximation around the time of horizon crossing.

\begin{figure}[t]
\begin{center}
 \begin{picture}(300,180)(0,0)
  \Line(0,170)(75,30)
  \Line(160,30)(160,170)
  \Photon(280,30)(280,170){2}{5}
  \Text(78,15)[]{\Large $N_0$}
  \Text(160,15)[]{\Large $N_i$}
  \Text(282,15)[]{\Large $N_f$}
  \SetColor{Red}
   \LongArrow(45,90)(157,90)
   \LongArrow(157,90)(45,90)
   \Text(100,105)[]{\Large $\Delta{N}_k$}
  \SetColor{Blue}
   \LongArrow(165,35)(275,35)
   \LongArrow(165,68)(275,68)
   \LongArrow(165,100)(275,100)
   \LongArrow(165,132)(275,132)
   \LongArrow(165,165)(275,165)
 \end{picture}
 \caption{Schematic display of how we use the $\delta N$ formalism. $N_0$ is the horizon crossing time of a mode, while $N_i$ is the fixed ``initial'' time, which we choose to be shortly after all modes of observational interest have crossed the horizon and hence is independent of $k$. $\Delta N_k$ is the number of $e$-foldings between these two times, which obviously depends on $k$. The final time, which we assume to be after all isocurvature modes have decayed and the curvature perturbation is conserved, is denoted by $N_f$. The two early times, $N_0$ and $N_i$ are defined on spatially flat hypersurfaces, while the final surface at $N_f$ is comoving~\cite{Sasaki:1995aw,deltaNetc}.}
 \label{fig:deltaN}
\end{center}
\end{figure}
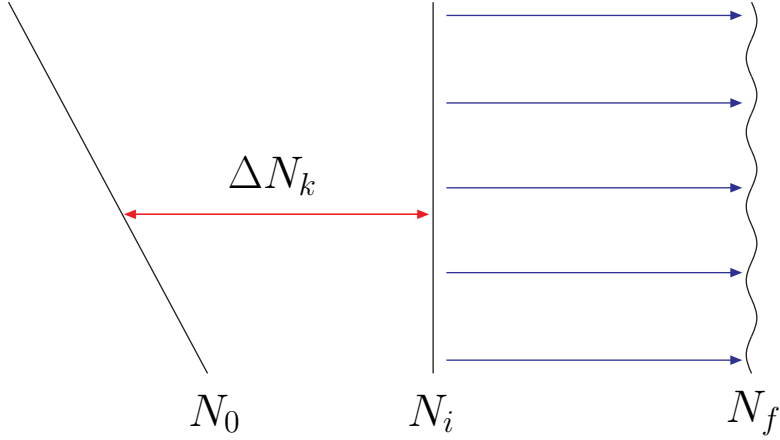

We take $t_i$ to be a fixed, pivot time shortly after all the modes of observational interest have crossed the horizon~\cite{Sasaki:1995aw}, while $t_0<t_i$ is the horizon crossing time of a certain mode $k=(aH)_0$, and hence $k$ dependent. Hence for each mode we find the number of $e$-folds between $t_0$ and $t_i$,
\begin{equation}
\Delta{N}_k = \log\left(\frac{a_i}{a_0}\right) \approx \log\left[\frac{(aH)_i}{k}\right] > 0 \, .
\end{equation}
This situation is depicted in Figure~\ref{fig:deltaN}.

An important point to notice, in order to describe the field fluctuation on the flat slices $\delta\phi^a \equiv \phi^a(t,\mathbi{x})-\phi^a_0(t)$ covariantly, is that in general it is coordinate dependent and thus is not covariant. To keep covariance, we realise that $\phi^a$ and $\phi^a_0$ are uniquely connected by a geodesic parametrised by $\lambda$ with respect to the field space metric $\gamma_{ab}$. Then, $\delta\phi^a$ can be expressed, with $\phi^a|_{\lambda=0}$ being identified with the background field, in terms of the {\em vector} $Q^a \equiv d\phi^a/d\lambda|_{\lambda=0}$ which lives in the tangent space at $\phi^a_0$. That is,
\begin{equation}
\phi^a - \phi^a_0 \equiv \delta\phi^a = Q^a - \frac{1}{2}\Gamma^a_{bc}Q^bQ^c + \cdots \, .
\end{equation}
Thus any tensor quantity written in terms of $Q^a$ is manifestly covariant by construction~\cite{Gong:2011uw}. Then, on very large scales where the space-time metric is that of unperturbed Friedmann-Robertson-Walker one, using the $e$-folds $N$ as the time variable the equation of motion for $Q^a$ is~\cite{Elliston:2012uy}
\begin{align}\label{Qevolution}
D_N Q^a & = w^a{}_b Q^b +\frac12 w^a{}_{bc}Q^bQ^c+\cdots \, , 
\\
w_{ab} & = u_{(a;b)} + \frac{R_{c(ab)d}}{3}\frac{\dot\phi^c_0}{H}\frac{\dot\phi^d_0}{H} \, ,
\\
w_{a(bc)} & = u_{(a;bc)} + \frac{1}{3} \left[ R_{(a|de|b;c)}\frac{\dot\phi^d_0}{H}\frac{\dot\phi^e_0}{H} - 4R_{a(bc)d}\frac{\dot\phi^d_0}{H} \right] \, ,
\\
u_a & = -\frac{V_{;a}}{3H^2} \, ,
\end{align}
where $D_N$ is a covariant derivative with respect to $N$, $R^a{}_{bcd}V^b \equiv V^a{}_{;cd}-V^a{}_{;dc}$ and a semicolon denotes a covariant derivative with respect to $\gamma_{ab}$. The use of the parentheses around field indices implies symmetrisation, and any terms between vertical bars are excluded from the symmetrisation. This implies that
\begin{equation}\label{Qevolution2}
Q^a(N_i=N_0+\Delta N) = Q^a(N_0) + \Delta{N}_k \left( w^a{}_b Q^b + \frac{1}{2} w^a{}_{bc} Q^bQ^c + \cdots \right) + \cdots \, . 
\end{equation}
Here as always in this text, we will work to leading order in $\Delta N$, as this is all we need in order to calculate the scale dependence of $\fnl$. Note that $\Delta N$ itself is order of a few (i.e.~of order the range of scales which we can observe, which corresponds to 5 - 10 $e$-foldings), but all $(\Delta N)^2$ terms will be multiplied by a second order in slow-roll term, unless $\nfnl$ is not small, in which case our formalism breaks down. This is analogous to the situation with the power spectrum~\cite{Nakamura:1996da,Gong:2002cx}, one may only treat the spectral index as being small and slowly varying if its running is further suppressed.

From the covariantised version of the $\delta N$ formalism \cite{Elliston:2012uy},
the comoving curvature perturbation $\calR$ is given by
\begin{equation}\label{deltaNformula}
\calR_k(t_f) = \delta{N} = N_a(t_i,t_f)Q_k^a(t_i) + \frac{1}{2}N_{ab}(t_i,t_f) \left[ Q^a(t_i)\star Q^b(t_i) \right]_k + \cdots \, ,
\end{equation}
where star denotes a convolution, and the final time $t_f$ is assumed to be after all the isocurvature perturbations have decayed and $\calR$ is conserved, so we will drop it from future expressions. In the case of a trivial field space metric, the $\delta N$ coefficients follow from a direct Taylor series expansion, $N_a=\partial N/ \partial \phi^a$ and so on~\cite{Lyth:2005fi}. We will not consider the case of isocurvature modes which persist until today in this paper. Note that based on our definition of $t_i$, the derivatives of $N$ are scale independent. See Appendix~B of~\cite{Byrnes:2010ft} for more details, and alternative ways to use the $\delta N$ formalism.

Now with the $k$-dependent factor manifest in (\ref{Qevolution2}) that comes from the evolution from the horizon crossing, which is different for each mode $k$, to the initial time for the $\delta{N}$ formalism, we first recapitulate the calculation of the power spectrum $\calP_\calR$ and the spectral index $n_\calR$.
For the power spectrum it is sufficient to work to linear order in $Q^a$ (in which case it reduces to $\delta\phi^a$). It is given by
\begin{align}
\left\langle \calR_\mathbi{k}(t_f)\calR_\mathbi{q}(t_f) \right\rangle = & (2\pi)^3\delta^{(3)}(\mathbi{k}+\mathbi{q}) \frac{2\pi^2}{k^3} \calP_\calR(k) = N_a(t_i)N_b(t_i) \left\langle Q_\mathbi{k}^a(t_i)Q_\mathbi{q}^b(t_i) \right\rangle \, .
\end{align}
Here, using (\ref{Qevolution2}) we can find
\begin{equation}\label{QQti-QQt0}
\left\langle Q_\mathbi{k}^a(t_i)Q_\mathbi{q}^b(t_i) \right\rangle = \left\langle Q_\mathbi{k}^a(t_0)Q_\mathbi{q}^b(t_0) \right\rangle + 2\Delta{N}_k w^a{}_c \left\langle Q_\mathbi{k}^b(t_0)Q_\mathbi{q}^c(t_0) \right\rangle \, ,
\end{equation}
where close to horizon crossing
\begin{equation}\label{QQ}
\left\langle Q_\mathbi{k}^aQ_\mathbi{q}^b \right\rangle = \frac{H^2}{2k^3} \delta^{(3)}(\mathbi{k}+\mathbi{q}) \left( \gamma^{ab} + \epsilon^{ab} \right) \, ,
\end{equation}
with $\epsilon^{ab}$ being first order in slow-roll and slowly varying~\cite{Sasaki:1995aw,Nakamura:1996da}, and can be found to arbitrary higher order in slow-roll using the Green's function approach~\cite{Gong:2002cx,greenfunctionsol}. Hence its derivative with respect to $\log k$ is second order in slow-roll. When taking the derivative of (\ref{QQ}) we need to take into account the scale dependence of $H^2(t_0)$ which gives rise to $-2\epsilon$ in the spectral index, but the covariant derivative of the metric is zero by definition, so this does not give rise to any scale dependence. Similarly for calculating the scale dependence, we may raise and lower indices using either $\gamma^{ab}(t_i)$ or $\gamma^{ab}(t_0)$ when working to the same level of precision. Finally notice that working to the same order, it is only necessary to specify the time dependence of background terms, and not those which are multiplied by $\Delta N$.

Hence we may straightforwardly calculate the spectral index from the above expressions, and the result is
\begin{equation}
n_\calR-1 = \frac{D \log\calP_\calR}{d\log k} = -2\epsilon - 2\frac{N_aN_bw^{ab}}{N_cN^c} \, ,
\end{equation}
which agrees with the known result~\cite{Sasaki:1995aw}.

Notice that this result, together with the observational constraint on the spectral index does imply that at least the particular combinations of $R_{abcd}$ which appear through $w^{ab}$ in the expression for the spectral index must be small, barring a chance cancellation between this and another term. This provides some justification for our assumption that terms involving derivatives of the field metric should be slow-roll suppressed.

\section{Bispectrum and the running of $\fnl$}
\label{sec:bispectrum}

The bispectrum of $\calR$ is defined by
\begin{equation}
\left\langle \calR_{\mathbi{k}_1}(t_f)\calR_{\mathbi{k}_2}(t_f)\calR_{\mathbi{k}_3}(t_f) \right\rangle \equiv (2\pi)^3\delta^{(3)}(\mathbi{k}_1+\mathbi{k}_2+\mathbi{k}_3)B_\calR(k_1,k_2,k_3) \, .
\end{equation}
We will make the usual assumption that the bispectrum due to the non-Gaussianity of the fields at horizon crossing~\cite{Elliston:2012uy} does not give rise to observable values of $\fnl$, which may be the case for DBI fields and non-Bunch-Davies vacuum. This is generally the case~\cite{noNGcrossing}, although there can be exceptions if the third derivative of the potential is large~\cite{largeNGcrossing}. There is no analogous proof in the case of a non-trivial field metric, but since we are assuming the Riemann tensor is slow-roll suppressed this assumption is likely to remain valid, at least in the vast majority of cases.

From (\ref{deltaNformula}), the three point function consists of two terms, 
\begin{align}\label{RRR}
\left\langle \calR_{\mathbi{k}_1}(t_f)\calR_{\mathbi{k}_2}(t_f)\calR_{\mathbi{k}_3}(t_f) \right\rangle = & N_aN_bN_c \left\langle Q^a_{\mathbi{k}_1}Q^b_{\mathbi{k}_2}Q^c_{\mathbi{k}_3} \right\rangle + \frac{1}{2} \left\{ N_{ab}N_cN_d \left\langle \left[ Q^a\star Q^b \right]_{\mathbi{k}_1}Q^c_{\mathbi{k}_2}Q^d_{\mathbi{k}_3} \right\rangle + \text{2 perm} \right\} \, .
\end{align}
In the above, all terms on the right hand side (RHS) should be evaluated at $t_i$. We first deal with the first term on the RHS of (\ref{RRR}). Using (\ref{Qevolution2}) and (\ref{QQ}), we can easily find
\begin{align}
& N_a(t_i)N_b(t_i)N_c(t_i) \left\langle Q^a_{\mathbi{k}_1}(t_i)Q^b_{\mathbi{k}_2}(t_i)Q^c_{\mathbi{k}_3}(t_i) \right\rangle
\nonumber\\
& = N_a(t_i)N_b(t_i)N_c(t_i) \left\{ \frac{1}{2}\Delta{N}_{k_1}w^a{}_{de} \left\langle \left[ Q^d(t_0)\star Q^e(t_0) \right]_{\mathbi{k}_1}Q^b_{\mathbi{k}_2}(t_0)Q^c_{\mathbi{k}_3}(t_0) \right\rangle + \text{2 perm} \right\}
\nonumber\\
& = (2\pi)^3 \delta^{(3)}(\mathbi{k}_1+\mathbi{k}_2+\mathbi{k}_3) N_a(t_i)N_b(t_i)N_c(t_i) \frac{H^4(t_0)}{4k_1^3k_2^3k_3^3} w^{abc} \left( k_1^3\Delta{N}_{k_1} + \text{2 perm} \right) \, .
\end{align}
For the second term of the RHS of (\ref{RRR}), it is useful to use (\ref{QQti-QQt0}) so it follows that
\begin{align}
& \frac{1}{2} N_{ab}(t_i)N_c(t_i)N_d(t_i) \left\langle \left[ Q^a(t_i)\star Q^b(t_i) \right]_{\mathbi{k}_1}Q^c_{\mathbi{k}_2}(t_i)Q^d_{\mathbi{k}_3}(t_i) \right\rangle
\nonumber\\
& = (2\pi)^3\delta^{(3)}(\mathbi{k}_1+\mathbi{k}_2+\mathbi{k}_3) N_{ab}(t_i)N_c(t_i)N_d(t_i) \frac{H^4(t_0)}{4k_1^3k_2^3} \left( \gamma^{ac}\gamma^{bd} + 2\Delta{N}_{k_1}w^{ac}\gamma^{bd} + 2\Delta{N}_{k_2}\gamma^{ac}w^{bd} \right) \, .
\end{align}

Putting both terms together, and specialising to an equilateral triangle, which is the only case we need to consider in order to calculate $\nfnl$ (for the justification see \cite{Byrnes:2010ft}), we find that $\fnl$ is given by
\begin{equation}\label{fNL}
\frac{6}{5}\fnl = \frac{N_{ab}N^a N^b}{\left(N_cN^c\right)^2} \left[ 1 + \Delta{N}_k \left( \frac{N_dN_eN_fw^{def}}{N_{gh}N^gN^h} + 4\frac{N_{de}N^dN_fw^{ef}}{N_{gh}N^gN^h} - 4\frac{N_dN_ew^{de}}{N_gN^g} \right) \right] \, .
\end{equation}
Therefore $\nfnl$ is simply the term multiplying $\Delta{N}_k$ in (\ref{fNL}), i.e.
\begin{equation}\label{nfnl}
\nfnl \equiv \frac{1}{\fnl}\frac{D\fnl}{d\log{k}} = -\frac{N_aN_bN_cw^{abc}}{N_{de}N^dN^e}+4w^{ab} \left(\frac{N_aN_b}{N_dN^d} -\frac{N_{ac}N_bN^c}{N_{de}N^dN^e}\right) \, .
\end{equation}
This general formula for the observable $\nfnl$ is the main result of this paper. The first term in (\ref{nfnl}) is due to the non-linearity of the field fluctuations which are generated between the time $t_0$ and $t_i$, i.e.~the correlator $\langle \delta\phi^3(t_i)\rangle$. Meanwhile, the remaining term of (\ref{nfnl}) arises from the effect of multiple fields contributing to $\calR$. Note that in the curved field space case, the effects of $\gamma^{ab}$ enter into both terms, which makes their separation less clear in this case than in (\ref{nfnl_flat}) below.

As can be read from the coefficients $w^{ab}$ and $w^{abc}$, naively $\nfnl = \calO(\epsilon)$. However since the derivative term does not appear in the formula for the spectral index, it could be larger and its value is model dependent. It is possible to have a large value of $\nfnl$~\cite{Kaiser:2012ak}, which should be a sharp prediction of such a model. It is estimated that for a fiducial value of $\fnl=50$, Planck can reach a sensitivity of $\Delta\nfnl=0.05$ depending on the sky coverage~\cite{Sefusatti:2009xu}.

As a check on our result (\ref{nfnl}), and to gain intuition, we demonstrate that our result reduces to the known expression in the case of a trivial field space metric. From~\cite{Byrnes:2010ft}, 
\begin{equation}\label{nfnl_flat}
n_{f_\mathrm{NL}}^\mathrm{(flat)} =  - \frac{F^{(2)}_{abc}N^aN^bN^c}{N_{d}N_{e}N^{de}}+ 4 \left( 2\sqrt{\epsilon_a\epsilon_b} - \eta_{ab}
\right) \left( \frac{N^aN^b}{N^dN_d} - \frac{N^aN^b_{c}N^c}{N^dN_{de}N^e} \right)\, ,
\end{equation}
where we have defined $\epsilon_a = (V_{,a}/V)^2/2$, $\eta_{ab} = V_{,ab}/V$ and
\begin{equation}
F^{(2)}_{abc} = {\sqrt{2}}\left(-4\sqrt{\epsilon_a\epsilon_b\epsilon_c}+\eta_{ab}\sqrt{\epsilon_c}
  +\eta_{bc}\sqrt{\epsilon_a}+\eta_{ca}\sqrt{\epsilon_b}-\frac{1}{\sqrt{2}}\frac{V_{,abc}}{3H^2}\right) \, . 
\end{equation}
Using $w_{ab}=-(V_{,a}/V)_{,b}=2\sqrt{\epsilon_a\epsilon_d} - \eta_{ad}$ and $w_{abc}=F^{(2)}_{abc}$ it follows that (\ref{nfnl}) reduces to (\ref{nfnl_flat}).

\section{Conclusions}
\label{sec:conclusions}

We have calculated a general formula for the scale dependence of $\fnl$, for the first time allowing for a curved field space metric. The non-trivial field space metric, which appears in the kinetic term of the scalar field Lagrangian, gives rise to new terms in $\nfnl$ depending on both the Riemann curvature tensor and its derivative with respect to the fields. The derivative terms do not appear in the analogous formula for the spectral index and hence could be significantly larger, since they are not constrained by observations showing that the spectral index is close to scale invariant.

Our work is motivated by the goal of connecting fundamental, high energy theories to the rapidly improving cosmological observations. Fundamental theories often predict the existence of large numbers of scalar fields, and of a curved field space metric. In 2013 the first relevant data from the Planck satellite will be released, and it could lead to a discovery of non-Gaussianity. In this case, constraints on not only its amplitude, but also its scale dependence will go a long way towards discriminating between the many models of the early universe, especially when combined with other measurements such as the spectral index and tensor-to-scalar ratio.

\subsection*{Acknowledgements}

We thank Joseph Elliston, Raquel Ribeiro, Misao Sasaki, David Seery and Reza Tavakol for useful discussions.
JG thanks the Aspen Center for Physics for hospitality, supported in part by the National Science Foundation under Grant No. 1066293, where part of this work was carried out. 
CB is supported by a Royal Society University Research Fellowship.
JG is supported in part by a Korean-CERN Fellowship.

\end{document}